\documentclass[%
    article,
    twocolumn,
preprintnumbers,
bibnotes,amsmath,amssymb,
aps,
pre,
floatfix,
]{revtex4-2}

\usepackage{graphicx}
\usepackage{dcolumn}
\usepackage{pifont}
\usepackage{tikz}
\usepackage{circuitikz}
\usepackage{mathtools}
\usepackage{bm}
\usepackage{amsmath,amssymb}
\usepackage{hyperref}
\usepackage{mathtools}
\usepackage{lipsum}

\newcommand{\bBranchFluxVector}{\mathbf{\Phi}_{\mathrm{b}}}
\newcommand{\bFluxVector}{\mathbf{\Phi}}
\newcommand{\bBranchFluxJJVector}{\mathbf{\Phi}_{\mathrm{b_J}}}
\newcommand{\bBranchFluxJJIndexedVector}{\Phi_{\mathrm{b_J}i}}
\newcommand{\bBranchFluxLVector}{\mathbf{\Phi}_{\mathrm{b_L}}}
\newcommand{\bIrrFluxVector}{\widetilde{\mathbf{\Phi}}}
\newcommand{\bAugIrrFluxVector}{\widetilde{\mathbf{\Phi}}_{+}}
\newcommand{\bExtFluxVector}{\mathbf{\Phi}_{\mathrm{x}}}
\newcommand{\bExtFluxIndexedVector}{\Phi_{\mathrm{x}i}}
\newcommand{\bAugFluxVector}{\widetilde{\mathbf{\Phi}}_+}

\newcommand{\bDotBranchFluxVector}{\Dot{\mathbf{\Phi}}_{\mathrm{b}}}
\newcommand{\bDotBranchFluxJJVector}{\Dot{\mathbf{\Phi}}_{\mathrm{b_J}}}

\newcommand{\bDotBranchIrrFluxVector}{\Dot{\widetilde{\mathbf{\Phi}}}_+}

\newcommand{\bD}{\mathbf{D}}
\newcommand{\bCeff}{\mathbf{C}_{\mathrm{eff}}}
\newcommand{\bM}{\mathbf{M}}
\newcommand{\bMplus}{\mathbf{M}_+}
\newcommand{\bMplusInv}{\mathbf{M}_+^{-1}}
\newcommand{\bMtrans}{\mathbf{M}^{\mathsf{T}}}
\newcommand{\bR}{\mathbf{R}}
\newcommand{\bC}{\mathbf{C}}
\newcommand{\bL}{\mathbf{L}}

\newcommand{\IrrFlux}{\widetilde{\varphi}}
\newcommand{\noVarIrrFlux}{\widetilde{\phi}}
\newcommand{\dotNoVarIrrFlux}{\Dot{\widetilde{\phi}}}

\newcommand{\noVarxFlux}[1]{\phi_{\mathrm{x}_{#1}}}

\newcommand{\DotNoVarIrrFluxWithIndex}[1]{\Dot{\widetilde{\phi}}_{#1}}

\newcommand{\Trans}{\mathsf{T}}

\newcommand{\mrm}[1]{\mathrm{#1}}

\newcommand{\IcPlus}{I_{\mrm{c}2+\mrm{c}1}}
\newcommand{\IcMinus}{I_{\mrm{c}2-\mrm{c}1}}

\newcommand{\vphiOnevphiTwoPlane}{$\varphi_1$\nobreakdash--$\varphi_2$~plane}

\newcommand{\arxiv}[1]{\href{http://arxiv.org/abs/#1}{\texttt{arXiv}:#1}}

\begin{document}

\preprint{\arxiv{2307.01926}}

\title{\texorpdfstring{Extracting Equations of Motion from Superconducting Circuits}{Extracting Equations of Motion from Superconducting Circuits}}

\author{Christian Z. Pratt}%
 \email{czpratt@ucdavis.edu}

\author{Kyle J. Ray}
    \email{kjray@ucdavis.edu}
 
\author{James P. Crutchfield}
    \email{chaos@ucdavis.edu}
\affiliation{%
Complexity Sciences Center and Department of Physics and Astronomy, University of California, Davis, One Shields Avenue, Davis, CA 95616
}%

\date{\today}


\begin{abstract}
Alternative computing paradigms open the door to exploiting recent innovations in computational hardware to probe the fundamental thermodynamic limits of information processing. One such paradigm employs superconducting quantum interference devices (SQUIDs) to execute classical computations. This, though, requires constructing sufficiently complex superconducting circuits that support a suite of useful information processing tasks and storage operations, as well as understanding these circuits' energetics. First-principle circuit design leads to prohibitive algebraic complications when deriving the effective equations of motion---complications that to date have precluded achieving these goals, let alone doing so efficiently. We circumvent these complications by (i) specializing our class of circuits and physical operating regimes, (ii) synthesizing existing derivation techniques to suit these specializations, and (iii) implementing solution-finding optimizations which facilitate physically interpreting circuit degrees of freedom that respect physically-grounded constraints. This leads to efficient and practical circuit prototyping, as well as accessing scalable circuit architectures. The analytical efficiency is demonstrated by reproducing the potential energy landscape generated by a SQUID. We then show how inductively coupling two SQUIDs produces a device that is capable of executing $2$-bit computations via its composite potential energy landscape. More generally, the synthesized methods detailed here provide a basis for constructing universal logic gates and investigating their thermodynamic performance.
\end{abstract}
\maketitle
\section{Introduction}
\label{sec:Introduction}

All computation is physical. To effect information processing, one approach entails a sequence of stochastic transformations that systematically manipulate a system's potential energy landscape \cite{Boyd_Crutchfield_2016,Ray_Boyd_Wimsatt_Crutchfield_2021}. Reliable computing, in particular, then requires stable memory states physically supported by a system's information-bearing degrees of freedom \cite{Landauer_1961}. Energy minima on the landscape provide this dynamical stability. Computation, then, consists of externally controlling the creation, destruction, and location of energy minima. From this perspective, a device's time-dependent potential energy surface guides the emergence of its computational capabilities. 

Exploring a superconducting circuit's ability to perform computational operations in this way involves understanding the device's energetics and subsequent
dynamical equations of motion \cite{Han_Lapointe_Lukens_1989,Han_1992_rf, Cantor_1996, Orlando_Mooij_Tian_1999,Muck_Chesca_Zhang_2001}. Success in using this approach to design candidate devices, though, requires rapidly determining if a given circuit construction is capable of carrying out computations. This requirement demands a framework that can efficiently derive a circuit's equations of motion. To accomplish this, we synthesize several previous approaches, specializing them to a class of circuits of practical interest. The result is a methodology for generating a readily-interpretable Lagrangian and associated equations of motion for a given circuit in terms of its classical degrees of freedom. 

The framework's success is demonstrated through two examples. First, we efficiently reproduce the Lagrangian of a SQUID \cite{Harada_Goto_Miyamoto_1987, Han_Lapointe_Lukens_1989, Han_1992_rf, Han_Lapointe_Lukens_1992, Harris_Johnson_Han_2008}. Then, we show how inductively coupling two SQUIDs produces a device that can execute a range of $2$-bit computations.

\section{Related Work}
\label{sec:RelatedWork}

The following synthesizes methods from Refs. \cite{Devoret_1995,Burkard_Koch_DiVincenzo_2004, You_Sauls_Koch_2019, Mariantoni_2020}. Its foundations build on Refs. \cite{Devoret_1995,Burkard_Koch_DiVincenzo_2004}, which introduced a network-theoretic approach to electrical circuit analysis and investigated circuits operating in the quantum regime. Reference \cite{You_Sauls_Koch_2019} introduced an elegant technique for multi-loop circuits to find \emph{irrotational} degrees of freedom. However, it considered only the circuit's quantum Hamiltonian for investigating time-dependent quantities, such as the transition probabilities between energy eigenstates. This departs from our goals. Moreover, to avoid cyclic coordinates in the equations of motion, Ref. \cite{You_Sauls_Koch_2019} restricted each circuit loop to have only a single inductor. The following eschews this restriction and, instead, finds optimal solutions for circuits containing more than one inductor in a loop by algebraically eliminating extra degrees of freedom.

Here, we use the resistive capacitive shunted junction (RCSJ) model for each Josephson junction (JJ). Due to this, the dissipative dynamics arising from finite-valued direct current (DC) resistances must be accounted for. To do this, we rely on Ref. \cite{Mariantoni_2020}'s method that uses the Rayleigh dissipation function \cite{Goldstein_Poole_Safko_2008} to model the circuit's resistive shunts.

Several alternative approaches are available to analyze circuit behaviors in the quantum regime. One common procedure employs number-phase quantization
\cite{Xiang-Guo_Ji-Suo_Yun_Hong-Yi_2008, Xiang-Guo_Ji-Suo_Bao-Long_2008}, which does not use a network-theoretic approach. Simulations of the quantum dynamics of similar circuits are detailed in Ref. \cite{Chitta_Zhao_Huang_Mondragon-Shem_Koch_2022}. This all noted, though the SQUIDs  employed here are often the basis for quantum computing devices, we concentrate on their behavior in the classical regime to understand their information-bearing degrees of freedom \cite{Han_Lapointe_Lukens_1992, Saira_Matheny_2020}. This also greatly facilitates follow-on investigations of their far-from-equilibrium thermodynamic performance.

Finally, a complementary approach to JJ circuit analysis considers the charge in a loop \cite{Ulrich_Hassler_2016}, as opposed to the magnetic flux. However, previous works \cite{Ray_Crutchfield_2022, Saira_Matheny_2020} revealed that varying magnetic flux provides a convenient circuit control method. Consequently, this grounds the following in a flux-focused interpretation of circuit behavior. A generalized approach to the techniques implemented in Ref. \cite{You_Sauls_Koch_2019} considers arbitrary circuit geometries and electromagnetic fields to construct a Hamiltonian \cite{Riwar_DiVincenzo_2022}. That said, analytical complications arising from this first-principles framework preclude rapidly and directly characterizing candidate circuit designs. The following circumvents such difficulties.

\section{Superconducting Circuit Analysis}
\label{sec:level3}

First, we obtain the equations of motion for a given circuit. Then, we show how to find coordinate transformations that produce readily interpretable equations of motion in Langevin form.

\subsection{Circuit Equations of Motion}\label{sec:CircuitEOM}

Following Ref. \cite{Devoret_1995}, we define a \textit{branch} to be a
particular circuit element, whose time dependent \emph{branch flux} is defined by:
\begin{align*}
\phi_b & = \phi_b(t) \\
       & \coloneqq \int_{-\infty}^{t} \mathrm{d}t' \; v_b(t') 
  ~.
\end{align*}
This is related to the branch voltage  $v_b(t)$, the instantaneous voltage across the circuit element, and the \emph{reduced branch flux} $\varphi_b = 2\pi \phi_b/ \Phi_0$, where
$\Phi_0$ is the flux quantum. 

Before proceeding, several assumptions need to be addressed. To begin, all
branches within a circuit correspond to either a Josephson junction (JJ) or an inductor.
Corresponding variables are subscripted with a $J$ or $L$, respectively.
All JJs are described by the RCSJ model \cite{Stewart_1968, McCumber_1968},
which is characterized by a critical current $I_{\mathrm{c}}$
\cite{Orlando_Mooij_Tian_1999}, capacitance $C_J$, and DC resistance $R$. Each
inductive branch is modeled by an inductance $L$ in parallel with a capacitance
$C_L$ satisfying the limit $C_L/C_J \approx 0$. We adopt $C_L$ as an auxiliary
variable in a fashion similar to Ref.  \cite{You_Sauls_Koch_2019}, in that the
limit is used at a particular step in the calculations, which is exemplified in Sections \ref{sec:vbrf} and \ref{sec:2vbrf}.

Suppose a circuit is constructed with $n$ JJs and $m$ inductors for a total of
$N = n + m$ branches. The \textit{branch flux} vector $\bBranchFluxVector
\coloneqq (\phi_{J_1}, \ldots,\phi_{J_n},\phi_{L_1},
\ldots,\phi_{L_m})^{\mathsf{T}}$ compactly represents all circuit branch fluxes.
When computing the potential and equations of motion, we refer to
the \textit{truncated branch flux} vectors $\bBranchFluxJJVector \coloneqq (\phi_{J_1},
\ldots, \phi_{J_n})^{\mathsf{T}}$ and $\bBranchFluxLVector \coloneqq (\phi_{L_1},
\ldots, \phi_{L_m})^{\mathsf{T}}$.
 
The energy stored in the capacitive components is \cite{Devoret_1995}:
\begin{align}
\label{eq:kinetic energy}
\mathcal{L}_T = \dfrac{1}{2} \bDotBranchFluxVector^\mathsf{T} \bC \bDotBranchFluxVector
	~,
\end{align}
where the overdot $\dot{}$ indicates a time derivative, and the \textit{capacitance} matrix is:
\begin{align*}
\bC \coloneqq \mathrm{diag}\;(C_{J_1},..., C_{J_n}, C_{L_1},...,C_{L_m})
  ~.
\end{align*}

Since we assume that all branches are either inductors  or JJs,
the energy stored in the inductive elements can be calculated using only 
$\bBranchFluxLVector$. The $m \times m$ \textit{inductance} matrix
$\bL$ denotes the circuit's linear inductances, with diagonal entries
corresponding to self-inductances $L_i$ and off-diagonal entries corresponding
to the mutual inductive coupling $-M_{ij}$ between $L_i$ and $L_{j\neq i}$. The energy
stored in the inductive components is given by \cite{Devoret_1995}:
\begin{align}
    \mathcal{L}_L = \dfrac{1}{2} \bBranchFluxLVector^{\mathsf{T}} \bL^{-1} \bBranchFluxLVector
\label{L potential energy}
	~.
\end{align}

Up to a constant, the JJ potential energy contribution is \cite{Devoret_1995}:
\begin{align}
\label{JJ potential energy}
\mathcal{L}_J = - \sum_{i=1}^n
	E_{i} \cos \left( \frac{2\pi}{\Phi_0} \bBranchFluxJJIndexedVector \right)
	~.
\end{align}
Here, $E_i =  I_{\mathrm{c}i}(\Phi_0/2\pi)$ is the Josephson energy of the $i$th JJ in a circuit, which is further characterized by its critical current $I_{\mathrm{c}i}$.

Equations (\ref{L potential energy})-(\ref{JJ potential energy})
together give the circuit's conservative potential energy $\mathcal{L}_V
\coloneqq \mathcal{L}_J + \mathcal{L}_L$. Given a physical circuit consisting
of inductors and JJs as described above, the circuit Lagrangian $\mathcal{L}
\coloneqq \mathcal{L}_T - \mathcal{L}_V$ is, up to a constant:
\begin{align}
\label{full Lagrangian}
\mathcal{L} = \dfrac{1}{2} \bDotBranchFluxVector^\mathsf{T} \bC \bDotBranchFluxVector 
	& - \dfrac{1}{2} \bBranchFluxLVector^\mathsf{T} \bL^{-1} \bBranchFluxLVector
	\nonumber \\
	& + \sum_{i = 1}^n E_{i}
	\cos \left( \frac{2\pi}{\Phi_0} \bBranchFluxJJIndexedVector \right)
	~.
\end{align}
The nonconservative dissipation from the finite JJ resistive shunts are taken
into account by the Rayleigh dissipation function $\mathcal{D}$, and further incorporated into the Euler-Lagrange equations of motion \cite{Goldstein_Poole_Safko_2008,Mariantoni_2020} in terms of a generalized coordinate $q_i$,  as:
\begin{align}
\label{FormalEOM}
    \dfrac{\mathrm{d}}{\mathrm{d}t} \dfrac{\partial \mathcal{L}}{\partial \Dot{q}_i} = \dfrac{\partial \mathcal{L}}{\partial q_i} -\dfrac{\partial \mathcal{D}}{\partial \Dot{q}_i} ~,
\end{align}
with
\begin{align}
\label{original dissipation function}
    \mathcal{D} \coloneqq \sum_{i=1}^{n} \dfrac{1}{2R_i} (\Dot{\phi}_{J_i})^2
	~.
\end{align}
$\mathcal{D}$ accounts for the dissipated power in each JJ branch due to its shunt resistance $R_i$ in terms of its branch flux $\phi_{J_i}$. To obtain this relation in matrix form, we first let:
\begin{equation}
    \bD = \mrm{diag}(R_{J_1}, ..., R_{J_n}) \nonumber ~.
\end{equation}
Then, the dissipation function reads:
\begin{align}\label{eq:dissipation function in matrix form}
    \mathcal{D} = \dfrac{1}{2} \bDotBranchFluxJJVector^{\mathsf{T}} \bD^{-1} \bDotBranchFluxJJVector ~.
\end{align} 
To conclude, we add the phenomenological contribution of the DC resistances' thermal noise
current to the equations of motion via:
\begin{align}
\label{eq:FormalEOM_thermal}
    \dfrac{\mathrm{d}}{\mathrm{d}t} \dfrac{\partial \mathcal{L}}{\partial \Dot{q}_i} = \dfrac{\partial \mathcal{L}}{\partial q_i} -\dfrac{\partial \mathcal{D}}{\partial \Dot{q}_i} + \eta_i(t) ~,
\end{align}
in which $\eta_i(t)$ is nonzero for the JJ branches only. One common method of modelling $\eta_i(t)$ is a Langevin treatment \cite{Han_Lapointe_Lukens_1992}, in which  we consider them to be statistically independent of each other, delta correlated over time, and determined by the fluctuation-dissipation theorem through the relation:
\begin{align*}
\langle \eta_i(t)\eta_j(t') \rangle = \frac{2k_B T}{R_i}\delta_{ij} \delta(t-t')
~.
\end{align*}

\subsection{Determining Optimal Coordinates}\label{sec:OptimalCoords}

Despite the fact that Eq. (\ref{FormalEOM}) marginally accommodates the
circuit's topology, it does not account for fluxoid quantization conditions
\cite{Yurke_Denker_1984, Devoret_1995}: These require that the sum of the branch
fluxes around any loop equals the external flux threading the loop. As a
result, while there may appear to be $N = n+m$ degrees of freedom in the
Lagrangian, there are only $N-F$ degrees of freedom in a circuit with $F$
independent loops---i.e., loops that contain no other loops---threaded by external fluxes.

In view of this, the \textit{external flux} vector $\bExtFluxVector
\coloneqq (\phi_{\mathrm{x}_1},...,\phi_{\mathrm{x}_F})^{\mathsf{T}}$ is defined
to cast fluxoid quantization into matrix form \cite{You_Sauls_Koch_2019}:
\begin{align*}
\bExtFluxVector = \mathbf{R} \bBranchFluxVector
~.
\end{align*}
The $F \times N$ matrix $\mathbf{R}$ is constructed in such a way that its
elements $R_{ij}$ satisfy the following criteria: Let $\mathsf{L}_i$
denote the $i$th loop threaded by the external flux $\bExtFluxIndexedVector$
that may contain branch flux $\phi_j$. Then:
\begin{align*}
    R_{ij} \coloneqq \begin{cases}
        +1 & \phi_{j} \in \mathsf{L}_i \text{ same orientation as } \bExtFluxIndexedVector\;, \\
        -1 & \phi_{j} \in \mathsf{L}_i \text{ 
        opposite orientation as } \bExtFluxIndexedVector\;, ~\text{and}\\
        0 & \phi_j \notin \mathsf{L}_i\;.
    \end{cases}
\end{align*}
Finally, the circuit's degrees of freedom are defined as $\bIrrFluxVector
\coloneqq (\Tilde{\phi}_1, \ldots,\Tilde{\phi}_{N-F})^{\mathsf{T}}$ \cite{You_Sauls_Koch_2019}. Generally,
these are a to-be-determined linear combination of the branch fluxes
represented by the $(N-F) \times N$ matrix $\bM$:
\begin{align*}
    \bIrrFluxVector = \mathbf{M}\bBranchFluxVector ~.
\end{align*}
Furthermore, due to fluxoid quantization, no more than $N-F$ degrees of freedom
in the circuit are expected. The quantization conditions are included by
employing the $N \times 1$ \textit{augmented flux} vector $\bAugFluxVector$ and the
$N \times N$ \textit{augmented} matrix $\mathbf{M}_+$ \cite{You_Sauls_Koch_2019}:
\begin{align*}
    \bAugFluxVector &\coloneqq 
    \begin{pmatrix}
    \bIrrFluxVector \\
    \bExtFluxVector
  \end{pmatrix} ~, \\[1pt]
  \bMplus &\coloneqq 
  \begin{pmatrix}
    \mathbf{M} \\
    \mathbf{R}
  \end{pmatrix}\;.
\end{align*}

Note that the branch flux vector and the augmented flux vector are directly
related to each other through $\bMplus$ by:
\begin{align}
\label{aug flux related to branch flux}
    \bAugFluxVector = \bMplus \bBranchFluxVector
	~.
\end{align}
With this, the circuit Lagrangian and associated equations of motion can be
written in terms of $\bAugIrrFluxVector$ by substituting
$\bBranchFluxVector=\bMplus^{-1} \bAugFluxVector$ into Eq. (\ref{full Lagrangian}) and Eq. (\ref{FormalEOM}), respectively.
Specifically, to find the circuit's Lagrangian in terms of $\bAugIrrFluxVector$,
$\bMplus$ must be invertible. Provided that the columns of $\bM$ are chosen to
be linearly independent of each other and of the columns of $\bR$, the
nonsingularity of $\bMplus$ is guaranteed.

However, ambiguity remains in defining the elements of $\bM$. Following Ref.
\cite{You_Sauls_Koch_2019}, these degrees of freedom are deemed
\textit{irrotational} by ensuring that they satisfy the following constraint:
\begin{align}
\label{eq:NullSpace}
    \bR \bC^{-1} \bM^{\mathsf{T}} = \mathbf{0}
	~.
\end{align}
This guarantees that the Lagrangian, when written in terms of $\bAugIrrFluxVector$,
does not depend on $\Dot{\mathbf{\Phi}}_\mathrm{x}$. Due to this,
$\bIrrFluxVector$ is referred to as the \emph{irrotational flux} vector, and $\bAugIrrFluxVector$ is the \emph{augmented irrotational flux} vector. In addition, Eq. (\ref{eq:NullSpace}) allows the equations of motion to be of Langevin form, further enabling thermodynamical analyses of the circuit's degrees of freedom---the subject of a sequel.

However, even after enforcing the irrotational constraint, there is still
additional freedom in defining $\bM$. To address this, we turn to the kinetic
energy term:
\begin{align}
\label{T}
    \mathcal{L}_{T} &= \dfrac{1}{2} \Dot{\bFluxVector}_{\mathrm{b}}^{\mathsf{T}} \bC \Dot{\bFluxVector}_{\mathrm{b}} \\
    &=  \dfrac{1}{2} \Dot{\bIrrFluxVector}_+^\mathsf{T} (\bMplus^{-1})^{\mathsf{T}} \bC \mathbf{M}_+^{-1} \Dot{\bIrrFluxVector}_+
  \nonumber \\
   &= \dfrac{1}{2} \Dot{\bIrrFluxVector}_+^{\mathsf{T}} \bCeff \Dot{\bIrrFluxVector}_+ \label{Ceff kinetic energy}\; ~,
\end{align}
in which $\bCeff$ is the \textit{effective capacitive} matrix. With Eq. (\ref{Ceff kinetic energy}) in mind, recall that the goal is to obtain
an easily interpretable Lagrangian and corresponding equations of motion for a
given circuit. A diagonal $\bCeff$ allows for a straightforward interpretation
of $\mathcal{L}_{T}$ as the kinetic energy in both the
$\bBranchFluxVector$ and the $\bIrrFluxVector$ bases. In other words, the task
is to find solutions of $\bM$ that yield a diagonal $\bCeff$. 

Analyzing a number of cases established a set of calculational guidelines that
result in a diagonal $\bCeff$ when solving for the components of $\bM$ through
Eq. (\ref{eq:NullSpace}). These aid in the task of finding optimal solutions in
the continuous family of possible solutions:
\begin{enumerate}
      \setlength{\topsep}{-8pt}
      \setlength{\itemsep}{-1pt}
      \setlength{\parsep}{-8pt}
\item The first $n$ rows of $\bM$ can each contain up to $n$ nonzero entries
	corresponding to the $n$ JJ coefficients of $\bM \bBranchFluxVector$, whose magnitudes are equivalent in this work. The other $m$ inductive elements of $\bM$---
	corresponding to the inductive coefficients in each of these rows---will either be zero or
	proportional to $C_L/C_J$; for the latter case, their magnitudes are arbitrarily chosen to satisfy Eq. (\ref{eq:NullSpace}). Once this constraint is satisfied, the limit $C_L/C_J \rightarrow 0$ is taken.
\item When $m-F>0$, the last $m-F$ rows of $\bM$ will each contain up to $m$ nonzero entries
    corresponding to the $m$ inductive flux coefficients of $\bM \bBranchFluxVector$: In each of these rows, the nonzero entries are equipped with magnitudes that satisfy Eq. (\ref{eq:NullSpace}). Meanwhile, each row's $n$ JJ coefficients will contain zero-valued entries.
\end{enumerate}
Importantly, linear independence between rows must be maintained when
implementing these conditions.

To briefly illustrate guideline (1), one possible realization is that in each
of the $n$ rows, every JJ coefficient takes on a nonzero value only once, while
all other JJ coefficients are zero. If each nonzero value is unity, this is
equivalent to there being no coordinate transformation between these branch and
irrotational flux coordinates.

Guideline (2) stems from a mismatch between the number of loops and inductors.
For example, if $m = 2$ and $F = 1$ such that $m-F = 1$---i.e., there is one loop that contains more
than one inductor---this requires setting all JJ coefficients to zero for one
solution of Eq. (\ref{eq:NullSpace}). This reflects the over-determination of the
additional inductor's behavior in the circuit. Consequently, one cyclic coordinate
will appear in the circuit Lagrangian: This can be eliminated through determining its
equation of motion and subsequently rewriting it in terms of noncyclic irrotational
degrees of freedom. Sections \ref{sec:vbrf} and \ref{sec:2vbrf} demonstrate this
procedure. Note that if $m - F = 0$, then guideline (2) does not apply. Additionally,
if $m - F < 0$, finding a diagonal $\bCeff$ is not possible.

Once the elements of $\bM$ are determined, the dynamical
degrees of freedom are interpreted as the irrotational degrees of freedom that
are not cyclic \cite{Goldstein_Poole_Safko_2008}. Numerically, there are $N - F
- (m - F) = n$, as there will be $N - F$ irrotational flux coordinates with $m -
F$ expected to be cyclic. For a multi-loop circuit ($F > 1$), a diagonal
$\bCeff$ is found only when there are no more JJs than there are irrotational
degrees of freedom. Equivalently, the number of inductors in a circuit
containing both JJs and inductors must satisfy $m \geq F$. These conditions can also be explained as the following: Each JJ
must be physically represented by at least one dynamical degree of freedom, and
there must be at least one inductor per independent circuit loop to capture the
circuit flux behavior. Below, we illustrate these conditions by example.

\section{Example Device Designs}
\label{sec:Examples}

The following demonstrates how to find the independent information-bearing degrees
of freedom via two examples: A SQUID and a device comprised of
two inductively coupled SQUIDs capable of implementing a range of $2$-bit computations.

\subsection{SQUID}
\label{sec:vbrf}

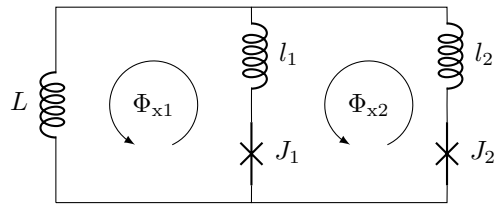
\begin{figure}[!htb]
  \begin{center}
    \begin{circuitikz}[scale=1.3] 

    \tikzstyle{every node}=[font=\normalsize]
    
      \draw (0,0)
      to[L=$L$, label distance=0.25cm] (0,2) 
      to[short] (2,2)
      to[L=$l_1$, label distance=0.25cm] (2,1) 
      to[barrier=$J_1$, label distance=-5cm] (2,0)
      to[short](0,0);
      
      \draw (2,2)
      to[short] (4,2)
      to[L=$l_2$, label distance=0.25cm] (4,1)
      to[barrier=$J_2$, label distance=-5cm] (4,0)
      to[short] (2,0);
      
      \draw[thin, ->, >=latex] (1,1)node{$\Phi_{\mathrm{x}1}$}  ++(-65:0.45) arc (-65:245:0.45);
      
       \draw[thin, ->, >=latex] (3.2,1)node{$\Phi_{\mathrm{x}2}$}  ++(-65:0.45) arc (-65:245:0.45);
      
    \end{circuitikz}
\caption{A SQUID with $N = 5$ and $F = 2$. Slight adjustments are made to the physical construction of the circuit compared to Ref. \cite{Han_1992_rf}.}
\label{fig:rF SQUID}
  \end{center}
\end{figure}

We first consider a circuit whose names and constructions span multiple use cases over a number of decades. Figure \ref{fig:rF SQUID} displays a circuit whose original name was the variable $\beta$ radio-frequency SQUID \cite{Han_Lapointe_Lukens_1989, Han_Lapointe_Lukens_1992, Han_1992_rf} and was later known as the compound Josephson junction radio-frequency SQUID \cite{Harris_Johnson_Han_2008, Harris_Lanting_2009}. The device's primary use cases involved investigating macroscopic quantum phenomenon, which deviates fundamentally from our goals. Applications that utilize the quantum flux parametron (QFP) \cite{Harada_Goto_Miyamoto_1987, Hosoya_Goto_1991, Takeuchi_2022} align more closely with our goals---employing superconducting devices for classical information processing---although the QFP construction and methods of operation differ from that of Refs. \cite{Han_Lapointe_Lukens_1989, Han_Lapointe_Lukens_1992, Han_1992_rf}. With this considered, we refer to the circuit in Fig. \ref{fig:rF SQUID} as a SQUID. 

Now, the goal is to reproduce the Lagrangian of the circuit shown in Fig. \ref{fig:rF SQUID}---whose design is detailed in Ref. \cite{Han_1992_rf}---using the methods detailed in Sections \ref{sec:CircuitEOM} and \ref{sec:OptimalCoords}. To accomplish this, we first begin writing out the flux vectors:
\begin{align}
\bBranchFluxVector & = (\phi_{J_1} \; \phi_{J_2} \; \phi_{L} \;
	\phi_{l_1} \; \phi_{l_2})^{\mathsf{T}} \label{eq:SQUID branch flux vector}~,\\
\bBranchFluxJJVector & = \left(\phi_{J_1} \;\; \phi_{J_2}
	\right)^{\mathsf{T}} ~,\\
\bBranchFluxLVector & = \left( \phi_{L} \;\; \phi_{l_1} \;\; \phi_{l_2}
	\right)^{\mathsf{T}} \label{eq:SQUID branch flux inductive vector}~, \\
\bAugIrrFluxVector & = (\noVarIrrFlux_1 \;\; \noVarIrrFlux_2 \;\;
	\noVarIrrFlux_3 \;\; \phi_{\mathrm{x}_1} \;\;
	\phi_{\mathrm{x}_2})^{\mathsf{T}} \label{eq:SQUID irrotational branch flux vector plus}
 ~.
\end{align}

With every branch orientation in Fig. \ref{fig:rF SQUID} pointing upwards, fluxoid quantization gives:
\begin{align*}
    \mathbf{R} = \begin{pmatrix}    
    1 & 0 & -1 & 1 & 0 \\
    -1 & 1 & 0 & -1 & 1
    \end{pmatrix}
	~,
\end{align*}
where each row's entries correspond to the column orientation of $(J_1, J_2,
L, l_1, l_2)$, and each row coincides with the external flux loops $(\phi_{\mathrm{x}_1}, \, \phi_{\mathrm{x}_2})$, respectively. Next, the capacitance matrix is written as:
\begin{align*}
\bC^{-1} = \mathrm{diag}(C_{J_1}^{-1}, C_{J_2}^{-1}, C_{L}^{-1}, C_{l_1}^{-1},
C_{l_2}^{-1})
  	~.
\end{align*}
To satisfy Eq. (\ref{eq:NullSpace}), let:
\begin{align*}
    \mathbf{M}^{\mathsf{T}} = \begin{pmatrix}
    M_{11} & M_{21} & M_{31} \\
    M_{12} & M_{22} & M_{32} \\
    M_{13} & M_{23} & M_{33} \\
    M_{14} & M_{24} & M_{34} \\
    M_{15} & M_{25} & M_{35} \\
    \end{pmatrix} ~.
\end{align*}
Then, with the assumption that $C_l \coloneqq C_{l_1} = C_{l_2} = C_L$ and $C_J
\coloneqq C_{J_1} = C_{J_2}$, each column of $\mathbf{M}^{\mathsf{T}}$
satisfies:
\begin{align}
    CM_{\mathrm{i}1} &= M_{\mathrm{i}3} - M_{\mathrm{i}4} \label{eq:rf m conditions 1} \\
    C(M_{\mathrm{i}2} - M_{\mathrm{i}1}) &= M_{\mathrm{i4}} - M_{\mathrm{i}5} \label{eq:rf m conditions 2} ~,
\end{align}
with $C \coloneqq C_l / C_J$ and $\mathrm{i} = 1,2,3$. 
From here, we use the guidelines described in Sec. \ref{sec:OptimalCoords} to obtain a diagonal $\bCeff$. This is achieved first via guideline (1) for the first $n=2$ rows of $\bM$ and guideline (2) for the last $m-F=1$ row of $\bM$. We then write a subset of the solution space of
Eqs. (\ref{eq:rf m conditions 1})-(\ref{eq:rf m conditions 2}) into the augmented matrix:
\begin{align}
\label{rF Mplus matrix}
    \bMplus & =   \begin{pmatrix}
    \mathbf{M} \\
    \mathbf{R}
  \end{pmatrix} \nonumber \\
  & =  \begin{pmatrix}
    1/2 & 1/2 & C/4 & -C/4 & -C/4 \\
    -1 & 1 & 0 & C & -C \\
    0 & 0 & 1 & 1 & 1 \\
    1 & 0 & -1 & 1 & 0 \\
    -1 & 1 & 0 & -1 & 1
    \end{pmatrix}
	~,
\end{align}
Consequently, we expect there to be $m-F=1$ cyclic irrotational degree of freedom once the circuit Lagrangian $\mathcal{L}$ is found. Next, taking the limit $C \rightarrow 0$ and then inverting $\bMplus$ yields:
\begin{align}
\label{rF MplusInv matrix}
    \bMplus^{-1} = \begin{pmatrix}
    1 & -1/2 & 0 & 0 & 0 \\
    1 & 1/2 & 0 & 0 & 0 \\
    2/3 & 0 & 1/3 & -2/3 & -1/3 \\
    -1/3 & 1/2 & 1/3 & 1/3 & -1/3 \\
    -1/3 & -1/2 & 1/3 & 1/3 & 2/3
    \end{pmatrix} ~,
\end{align}
which aids in computing the effective capacitive matrix from Eq. (\ref{Ceff kinetic energy}) as:
\begin{align*}
    \bCeff = \begin{pmatrix}
    2C_J & 0 & 0 & 0 & 0 \\
    0 & C_J/2 & 0 & 0 & 0 \\
    0 & 0 & 0 & 0 & 0 \\
    0 & 0 & 0 & 0 & 0 \\
    0 & 0 & 0 & 0 & 0 \\
    \end{pmatrix} ~,
\end{align*}
which is diagonal as expected due to following guidelines (1) and (2).

As there are no mutual inductance couplings, the inductance matrix is:
\begin{align*}
    \bL = \begin{pmatrix}
        L & 0 & 0 \\
        0 & l_1 & 0 \\
        0 & 0 & l_2
    \end{pmatrix} ~.
\end{align*}

Recalling Eq. (\ref{aug flux related to branch flux}) and writing the circuit
Lagrangian from Eq. (\ref{full Lagrangian}) in terms of irrotational
branch fluxes, produces:
\begin{align}
    \mathcal{L} &= \dfrac{C_J}{2}\left( 2\dotNoVarIrrFlux_1^{\; 2} +  \dfrac{1}{2} \dotNoVarIrrFlux_2^{\; 2} \right) 
    \nonumber \\
    &~~-\dfrac{1}{9L} \left(2\noVarIrrFlux_1 + \noVarIrrFlux_3 - 2\noVarxFlux{1} - \noVarxFlux{2} \right)^2 \nonumber \\
    &~~-\dfrac{1}{9l_1} \left( -\noVarIrrFlux_1 + \frac{3}{2}\noVarIrrFlux_2 + \noVarIrrFlux_3 + \noVarxFlux{1} - \noVarxFlux{2} \right)^2 \nonumber \\
    &~~-\dfrac{1}{9l_2} \left(-\noVarIrrFlux_1 - \frac{3}{2}\noVarIrrFlux_2 + \noVarIrrFlux_3 + \noVarxFlux{1} + 2\noVarxFlux{2} \right)^2 \nonumber \\
    &~~+E_{2+1} \cos\IrrFlux_1 \cos\dfrac{\IrrFlux_2}{2} - E_{2-1} \sin \IrrFlux_1 \sin \dfrac{\IrrFlux_2}{2} \label{eq: 1 vbeta-rF SQUID} ~,
\end{align}
where $E_{2\pm1} = E_{J_2} \pm E_{J_1}$.

The Lagrangian is independent of $\dotNoVarIrrFlux_3$ indicating that it is, as expected, a cyclic
degree of freedom: It can be eliminated by computing the Euler-Lagrange equation of motion, and utilizing appropriate relations from Ref. \cite{Han_1992_rf}---specifically $l \coloneqq l_1 = l_2 \ll L$. From this, we find that  $\noVarIrrFlux_3
= \noVarIrrFlux_1 - \noVarxFlux{1} -
\noVarxFlux{2}/2$, and further substitute this into $\mathcal{L}$. A map between the circuit flux variables in Eq. (\ref{eq: 1 vbeta-rF SQUID}) and those from Ref. \cite{Han_1992_rf} can then be identified as:
\begin{align}
    \noVarIrrFlux_{1} &= \phi \label{eq: rf flux relation}~,\\
    \noVarIrrFlux_{2} &= \phi_{\mrm{dc}}\label{eq: dc flux relation} ~,\\
    \noVarxFlux{1} &= \phi_{\mrm{x}} - \dfrac{1}{2} \phi_{\mrm{xdc}} ~,
	~\text{and}\\
    \noVarxFlux{2} &= \phi_{\mrm{xdc}} ~.
\end{align}
Making these substitutions into Eq. (\ref{eq: 1 vbeta-rF SQUID}) yields a Lagrangian
$\mathcal{L}$ that matches that of Ref. \cite{Han_1992_rf}
with the preceding variable substitutions:
\begin{align}
\label{eq:VBRF_squid}
    \mathcal{L} &= \mathcal{L}_T - \mathcal{L}_{\mathrm{SQUID}} \\
    &= \dfrac{C_J}{2}\left( 2\Dot{\phi}^2 + \dfrac{1}{2}\Dot{\phi}_{\mathrm{dc}}^2\right)\nonumber \\
    & \qquad - \dfrac{1}{2L}(\phi - \phi_{\mathrm{x}})^2 - \dfrac{1}{4l}(\phi_{\mathrm{dc}} - \phi_{\mathrm{xdc}})^2 \nonumber \\[2pt]
    & \qquad + E_{2+1} \cos\varphi \cos\dfrac{\varphi_{\mathrm{dc}}}{2} - E_{2-1} \sin \varphi \sin \dfrac{\varphi_{\mathrm{dc}}}{2} \nonumber
  ~.
\end{align}
The goal is to now obtain the full Langevin equation of motion in Eq. (\ref{eq:FormalEOM_thermal}) for the circuit in Fig. \ref{fig:rF SQUID}. Since we wrote the circuit Lagrangian in Eq. (\ref{eq:VBRF_squid}) in terms of the coordinates $\phi$ and $\phi_{dc}$, we now need to transform the dissipative contribution in Eq. (\ref{eq:dissipation function in matrix form}) into the irrotational flux basis. Assuming both JJ shunts have a DC resistance of $R$, we have $\bD^{-1} = \mathrm{diag}(1/R, \; 1/R) ~.$ This leads to:
\begin{align}
    \mathcal{D} &= \dfrac{1}{2} \bDotBranchFluxJJVector^{\Trans} \bD^{-1} \bDotBranchFluxJJVector \nonumber \\[0.5pt]
    &= \dfrac{1}{2R} \left(\Dot{\phi}_{J_1}^2 + \Dot{\phi}_{J_2}^2  \right) ~.
\end{align}

Now, we use $\bMplusInv$ from Eq. \eqref{rF MplusInv matrix} to write the branch fluxes in terms of irrotational coordinates with the relation $\bDotBranchFluxJJVector = \bMplusInv \bDotBranchIrrFluxVector$ from Eq. \eqref{aug flux related to branch flux}:
\begin{align}
    \Dot{\phi}_{J_{1,2}} &= \DotNoVarIrrFluxWithIndex{1} \mp \dfrac{1}{2} \DotNoVarIrrFluxWithIndex{2} = \Dot{\phi} \mp \dfrac{1}{2} \Dot{\phi}_{\mrm{dc}} ~,
\end{align}
where the last equality was obtained using Eqs. \eqref{eq: rf flux relation} and \eqref{eq: dc flux relation}. Further substituting this into the dissipation function yields:
\begin{equation}\label{eq:Han dissipation function}
    \mathcal{D} = \dfrac{1}{R} \left(\Dot{\phi}^2 + \dfrac{1}{4}\phi_{\mathrm{dc}}^2 \right) ~.
\end{equation}

By using Eq. (\ref{eq:FormalEOM_thermal}), we can write the Langevin equation of motion in terms of the dynamical variables $\phi$ and $\phi_{\mathrm{dc}}$:
\begin{align}
    2C_J \Ddot{\phi} &= -\dfrac{1}{L}(\phi - \phi_{\mathrm{x}}) - \IcPlus\sin \varphi \cos \dfrac{\varphi_{\mrm{dc}}}{2} \nonumber\\[0.5pt]
    &\quad-\IcMinus \cos \varphi \sin \dfrac{\varphi_{\mrm{dc}}}{2} - \dfrac{2}{R}\Dot{\phi} + \eta (t) ~, \label{eq:EOM rf loop}\\[0.5pt]
    \dfrac{C_J}{2}\Ddot{\phi}_{\mrm{dc}} &= -\dfrac{1}{2l}(\phi_{\mrm{dc}} - \phi_{\mrm{xdc}}) - \IcPlus \cos \varphi \sin \dfrac{\varphi_{\mrm{dc}}}{2} \nonumber\\[0.5pt]
    &\quad-\IcMinus \sin \varphi \cos \dfrac{\varphi_{\mrm{dc}}}{2} - \dfrac{1}{2R} \Dot{\phi}_{\mrm{dc}} + \eta_{\mrm{dc}}(t) . \label{eq:EOM dc loop}
\end{align}
Eqs. (\ref{eq:EOM rf loop}) and (\ref{eq:EOM dc loop}) can then be employed to investigate the thermodynamic efficiency of the device in Fig. \ref{fig:rF SQUID}.

\subsection{Inductively Coupled SQUIDs}
\label{sec:2vbrf}

\begin{figure}[!t]
    \begin{center}
     \begin{circuitikz}[scale=1.25]
    \tikzstyle{every node}=[font=\normalsize]

    \draw (0,0)
    to[L=$L_1$, label distance=0.0cm] (3,0)
    to[short] (3,2)
    to[barrier=$J_1$, label distance=-5cm] (1.5,2)
    to[L=$l_1$, label distance=0.0cm] (0,2)
    to[short] (0,0);

    \draw (3,2)
    to[short] (3,4)
    to[barrier=$J_2$, label distance=-5cm] (1.5,4)
    to[L=$l_2$] (0,4)
    to[short] (0,2);

    \draw (3,-0.6)
    to[L=$L_2$] (0,-0.6)
    to[short] (0,-2.6)
    to[L=$l_3$] (1.5, -2.6)
    
    to[barrier] (3,-2.6) 
    to[short] (3,-0.6);

    \draw (0,-2.6)
    to[short] (0,-4.6)
    to[L=$l_4$] (1.5,-4.6)
    to[barrier] (3,-4.6) 
    to[short] (3,-2.6);
    
      
    \draw[thin, ->, >=latex] (1.5,1.05)node{$\mathrm{\phi_{1\mathrm{x}}}$}  ++(-65:0.50) arc (-65:245:0.45);
      
    \draw[thin, ->, >=latex] (1.5,2.8)node{$\mathrm{\phi_{1\mathrm{xdc}}}$}  ++(-65:0.50) arc (-65:245:0.47);

    \draw[thin, ->, >=latex] (1.5,-1.8)node{$\mathrm{\phi_{2\mathrm{x}}}$}  ++(-65:0.50) arc (-65:245:0.45);

    \draw[thin, ->, >=latex] (1.5,-3.55)node{$\mathrm{\phi_{2\mathrm{xdc}}}$}  ++(-65:0.50) arc (-65:245:0.47);

    \draw[-] (1.1,-0.25) -- (1.9,-0.25) node{};
    \draw[-] (1.1,-0.35) -- (1.9,-0.35) node{};

    \node at (2.3,-0.3) {$M$};

    \node at (2.2, -2.2) {$J_3$};

    \node at (2.2, -4.2) {$J_4$};

    \end{circuitikz}
\caption{Two SQUIDs inductively coupled via $M$: A superconducting
	device that supports $2$-bit classical computations through the manipulation of its potential energy landscape.
	}
\label{fig:two bit}
\end{center}
\end{figure}
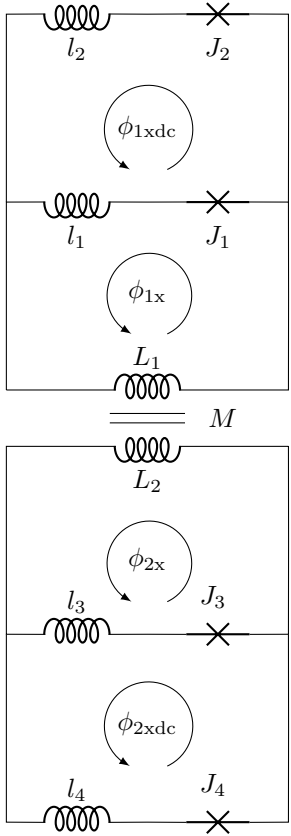

For our final example, consider inductively coupling two SQUIDs through $L_1$ and $L_2$ via the mutual inductance coupling constant $M \coloneqq M_{12} = M_{21}$, shown in Fig. \ref{fig:two bit}. Using the methods described in Secs. \ref{sec:CircuitEOM} and \ref{sec:OptimalCoords}, as well as the results from Sec. \ref{sec:vbrf}, allows rapidly deriving its potential. After this, we  discuss how this controllable potential performs $2$-bit computations.

The choice of branch orientation for the circuit in Fig. \ref{fig:two bit} is represented by:
\begin{align*}
    \bR = \begin{pmatrix}
        1 & 0 & 0 & 0 & -1 & 0 & 1 & 0 & 0 & 0 \\
        -1 & 1 & 0 & 0 & 0 & 0 & -1 & 1 & 0 & 0 \\
        0 & 0 & 1 & 0 & 0 & -1 & 0 & 0 & 1 & 0 \\
        0 & 0 & -1 & 1 & 0 & 0 & 0 & 0 & -1 & 1 
    \end{pmatrix} ~,
\end{align*}
in which each row's elements coincide with the column orientation $(J_1, J_2, J_3, J_4, L_1, L_2, l_1, l_2, l_3, l_4)$, each row corresponds to the external flux loop $(\phi_{1\mathrm{x}}, \, \phi_{1\mathrm{xdc}}, \, \phi_{2\mathrm{x}}, \, \phi_{2\mathrm{xdc}})$, and each branch orientation of the upper [lower] SQUID points left [right]. Using the irrotational constraint $\bR \bC^{-1} \bMtrans = \mathbf{0}$, we find
that the elements of $\bM$ need to satisfy:
\begin{align*}
    CM_{\mrm{i}1} &= M_{\mrm{i}5} - M_{\mrm{i}7} \\
    C(M_{\mrm{i}2} - M_{\mrm{i}1}) &= M_{\mrm{i}7} - M_{\mrm{i}8} \\
    CM_{\mrm{i}3} &= M_{\mrm{i}6} - M_{\mrm{i}9} \\
    C(M_{\mrm{i}4} - M_{\mrm{i}3}) &= M_{\mrm{i}9} - M_{\mrm{i}10} ~.
\end{align*}
Taking a lesson from the single SQUID case, and after taking $C
\rightarrow 0$, our choice of $\bM$ and $\bR$ leads to:
\begin{align*}
    \bMplus = \begin{pmatrix}
        1/2 & 1/2 & 0 & 0 & 0 & 0 & 0 & 0 & 0 & 0 \\
        -1 & 1 & 0 & 0 & 0 & 0 & 0 & 0 & 0 & 0 \\
        0 & 0 & 1/2 & 1/2 & 0 & 0 & 0 & 0 & 0 & 0 \\
        0 & 0 & -1 & 1 & 0 & 0 & 0 & 0 & 0 & 0 \\
        0 & 0 & 0 & 0 & 1 & 0 & 1 & 1 & 0 & 0 \\
        0 & 0 & 0 & 0 & 0 & 1 & 0 & 0 & 1 & 1 \\
        & & & & \bR
    \end{pmatrix}
	~,
\end{align*}
whose inverse is:
\begin{widetext}
\begin{align*}
    \bMplusInv = \begin{pmatrix}
        1 & -1/2 & 0 & 0 & 0 & 0 & 0 & 0 & 0 & 0 \\
        1 & 1/2 & 0 & 0 & 0 & 0 & 0 & 0 & 0 & 0 \\
        0 & 0 & 1 & -1/2 & 0 & 0 & 0 & 0 & 0 & 0 \\
         0 & 0 & 1 & 1/2 & 0 & 0 & 0 & 0 & 0 & 0 \\
         2/3 & 0 & 0 & 0 & 1/3 & 0 & -2/3 & -1/3 & 0 & 0 \\
         0 & 0 & 2/3 & 0 & 0 & 1/3 & 0 & 0 & -2/3 & -1/3 \\
         -1/3 & 1/2 & 0 & 0 & 1/3 & 0 & 1/3 & -1/3 & 0 & 0 \\
         -1/3 & -1/2 & 0 & 0 & 1/3 & 0 & 1/3 & 2/3 & 0 & 0 \\
         0 & 0 & -1/3 & 1/2 & 0 & 1/3 & 0 & 0 & 1/3 & -1/3 \\
         0 & 0 & -1/3 & -1/2 & 0 & 1/3 & 0 & 0 & 1/3 & 2/3\\
    \end{pmatrix}
	~.
\end{align*}
\end{widetext}
We then eliminate the cyclic degrees of freedom $\noVarIrrFlux_5$ and
$\noVarIrrFlux_6$. Following the single SQUID case detailed in Sec. \ref{sec:vbrf}, the map between our flux variables and those of Ref. \cite{Han_1992_rf} is:
\begin{align*}
    \noVarIrrFlux_{i} &= \phi_j ~,\\
    \noVarIrrFlux_{i+\mrm{1}} &= \phi_{j\mrm{dc}} ~,\\
    \noVarxFlux{i} &= \phi_{j\mrm{x}} - \dfrac{1}{2} \phi_{j\mrm{xdc}} ~, ~\text{and} \\
    \noVarxFlux{i+1} &= \phi_{j\mrm{xdc}}
    ~.
\end{align*}
Here, the index $i$ corresponds to either the $i$th dynamical degree of freedom or the $i$th external flux, while the index $j$ aligns with the flux in the $j$th SQUID, for which $i = 1,3$ and $j = 1,2$, respectively. 

Next, the inductive contribution to the potential, when taking $L \coloneqq L_1
= L_2$ and $l \coloneqq l_1 = l_2 = l_3 = l_4$, is found by first writing:
\begin{align*}
    \bL = \begin{pmatrix}
        L & -M & 0 & 0 & 0 & 0 \\
        -M & L & 0 & 0 & 0 & 0 \\
        0 & 0 & l & 0 & 0 & 0 \\
        0 & 0 & 0 & l & 0 & 0 \\
        0 & 0 & 0 & 0 & l & 0 \\
        0 & 0 & 0 & 0 & 0 & l
    \end{pmatrix} ~.
\end{align*}
Then, subsequently taking its inverse gives:
\begin{align*}
    \bL^{-1} = \begin{pmatrix}
        1/L_\alpha & \mu/L_\alpha & 0 & 0 & 0 & 0 \\
        \mu/L_\alpha & 1/L_\alpha & 0 & 0 & 0 & 0 \\
        0 & 0 & 1/l & 0 & 0 & 0 \\
        0 & 0 & 0 & 1/l & 0 & 0 \\
        0 & 0 & 0 & 0 & 1/l & 0 \\
        0 & 0 & 0 & 0 & 0 & 1/l
    \end{pmatrix} ~,
\end{align*} 
where $L_\alpha = \alpha L$, $\alpha = 1 - \mu^2$, and $\mu = M / L$.

With this, the potential is then:
\begin{align}
    \mathcal{L}_{V} &= -E_{2+1}\cos\varphi_1 \cos\dfrac{\varphi_{1\mathrm{dc}}}{2} + E_{2-1}\sin\varphi_1 \sin \dfrac{\varphi_{1\mathrm{dc}}}{2} \nonumber \\[2pt]
    &~~ -E_{4+3}\cos\varphi_2 \cos \dfrac{\varphi_{2\mathrm{dc}}}{2} + E_{4-3}\sin\varphi_2 \sin \dfrac{\varphi_{2\mathrm{dc}}}{2} \nonumber \\
     &~~ + \dfrac{1}{4l} (\phi_{1\mathrm{dc}} - \phi_{1\mathrm{xdc}})^2 + \dfrac{1}{4l} (\phi_{2\mathrm{dc}} - \phi_{2\mathrm{xdc}})^2 \nonumber \\
    &~~ + \dfrac{1}{2L_\alpha}(\phi_1 - \phi_{1\mathrm{x}})^2 + \dfrac{1}{2L_\alpha}(\phi_2 - \phi_{2\mathrm{x}})^2 \nonumber \\[2pt]
    &~~ + \dfrac{\mu}{L_\alpha}(\phi_1 - \phi_{1\mathrm{x}})(\phi_2 - \phi_{2\mathrm{x}}) ~. \label{the alpha potential} 
\end{align}

If we assume small coupling by keeping only linear terms in $\mu$, then
$L_\alpha^{-1} \to L^{-1}$, resulting in Eq. (\ref{the alpha potential})
simplifying to a sum of two SQUIDs potential contributions and
a mutual inductance coupling $\mathcal{L}_{\mathrm{M.I.}}$:
\begin{align}
\label{eq:simplified potential}
    \mathcal{L}_V = \mathcal{L}_{\mathrm{SQUID}\text{-}1} + \mathcal{L}_{\mathrm{SQUID}\text{-}2} + \mathcal{L}_{\mathrm{M.I.}} ~,
\end{align}
in which $\mathcal{L}_{\mathrm{M.I.}} = \mu(\phi_1 - \phi_{1\mathrm{x}})(\phi_2 - \phi_{2\mathrm{x}})/L$. This Lagrangian can be used in much the same way as that in Eq. \eqref{eq:VBRF_squid} to obtain the full equations of motion through Eq. (\ref{eq:FormalEOM_thermal}).

Follow-up efforts aim to use Eq. (\ref{eq:simplified potential}) to perform computations. One way of accomplishing this goal is to reduce Eq. (\ref{eq:simplified potential}) to an effective two-dimensional surface of interest: This surface can contain energy minima that are separated by energy barriers, whose heights exceed the thermal energy $k_B T$. To achieve this, we make the following choices: First, $E_{2+1} = E_{4+3} = 1.05 \times 10^{-21} \; \text{J}$ \cite{Ozfidan_2020}, while $E_{2-1} = E_{4-3} = 0$. The latter assumption instantiates symmetry into the potential. Next, we assume $l \ll L$: Consequently, any modifications of the circuit parameter $\phi_{i\mathrm{xdc}}$ will be reflected in $\phi_{i\mathrm{dc}}$ on small enough timescales that we assume $\phi_{i\mathrm{dc}} = \phi_{i\mathrm{xdc}}$. Finally, $L = 230$ pH \cite{Ozfidan_2020}. 

\begin{figure}[!htb]
\centering
\includegraphics[width=0.9\linewidth]{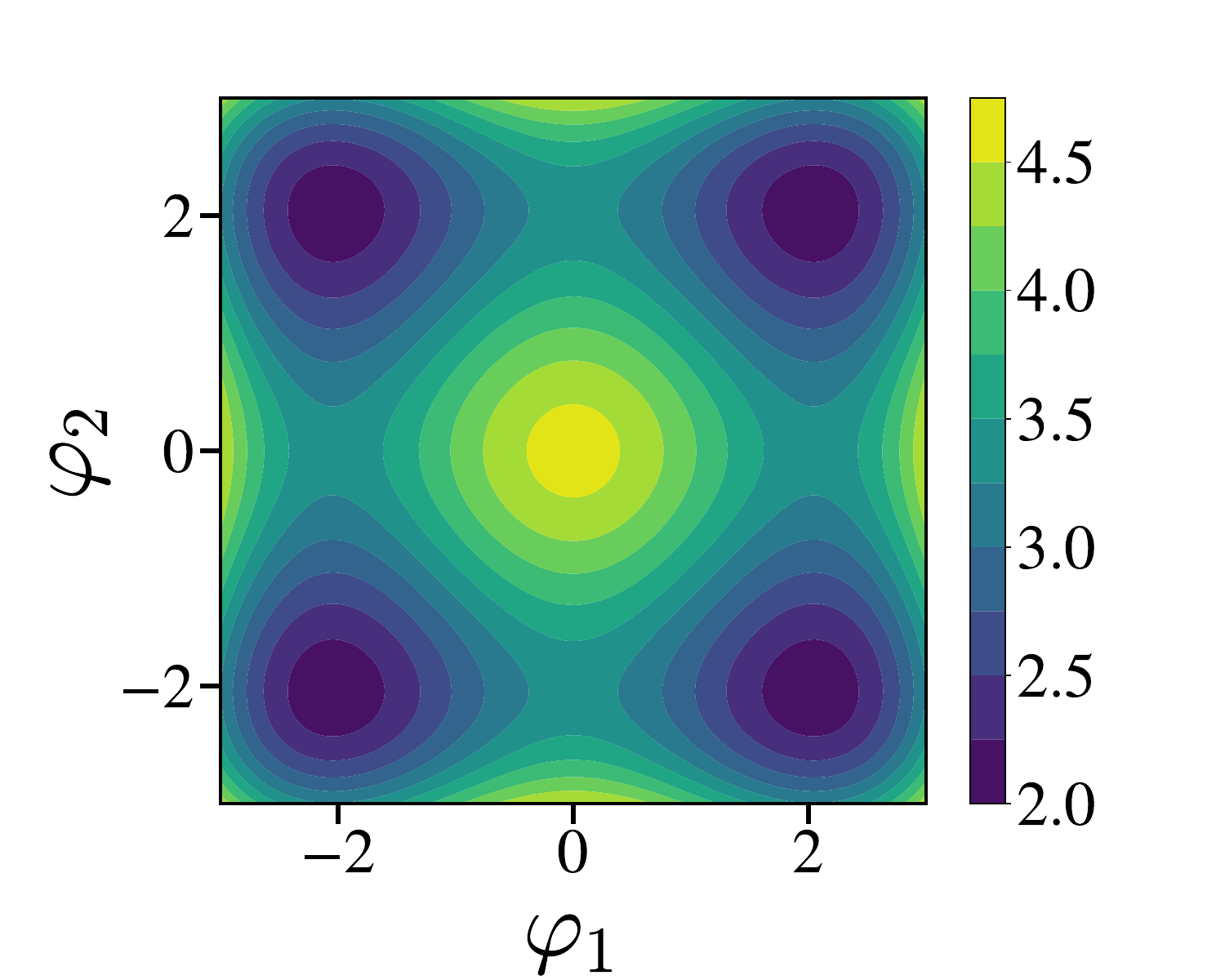}
\caption{Effective energy landscape generated by Eq. (\ref{eq:simplified potential}) after making simplifications and assumptions that introduce symmetry into the potential. Additionally, all other external circuit parameters take on zero value. Instantiating each region surrounding an energy minimum as a computational memory state---$00$, $10$, $01$, and $11$---permits information storage. Information processing is accomplished by way of a control protocol that employs the mutual inductance and external flux parameters to transform the landscape.}
\label{fig:comp potential}
\end{figure}

After implementing these selections, Eq. (\ref{eq:simplified potential}) is reduced to the effective two-dimensional potential energy surface in the \vphiOnevphiTwoPlane $\;$shown in Fig. \ref{fig:comp potential}. Note that this potential currently has neutral external parameter values, i.e. $\phi_{1\mathrm{xdc}} = \phi_{2\mathrm{xdc}} = 0$ and $\mu = 0$. This potential contains four stable energy minima that can each be assigned a computational memory state---$00$, $01$, $10$, and $11$. Taking advantage of the metastable regions near each minimum, we can store information. By varying the values of $M$, $\phi_{i\mathrm{x}}$, and
$\phi_{i\mathrm{xdc}}$ for which $i = 1, 2$, we can process that information---using the dynamics of
the Euler-Lagrange equation of motion via Eq. (\ref{eq:FormalEOM_thermal}) to implement $2$-bit logic gates. Note that while we considered $M$ to be a tunable coupling constant, the details of its construction---a SQUID coupler---are detailed in Refs. \cite{Brink_Berkley_Yalowsky_2005, Harris_Berkley_Johnson_2007, Harris_Lanting_2009}. The coupler's equations of motion could be accounted for within the complete device construction if its dynamics become important in future investigations.

\section{Conclusion}
\label{sec:Conclusion}

We introduced a method that enables exploring the classical informational processing properties of a candidate superconducting circuit through understanding the circuit's energetics and subsequent dynamics. Through examples, we demonstrated the analytical efficiency of the method by reproducing the equations of motion for a SQUID that implements single bit computations \cite{Saira_Matheny_2020}, as well as a more complicated device that supports $2$-bit computations. 

This is the first effort in a series on physically-realizable classical
computing. In point of fact, the coupled SQUIDs shown in Fig.
\ref{fig:two bit} also supports the information processing behavior exhibited
by a Szilard engine \cite{szilard1925ausdehnung, Szilard_1964,
Boyd_Crutchfield_2016}. Follow-on efforts explore the dynamical and thermodynamic properties
of these circuits and implement universal gates---e.g., NAND and Fredkin.

\section{Acknowledgments}
\label{sec:Acknowledgements}

The authors thank Camron Blackburn, Scott Habermehl, Jukka Pekola, Paul Riechers, Kuen Wai Tang, and Greg Wimsatt for helpful comments and discussions, as well as the
Telluride Science Research Center for its hospitality during visits and the
participants of the Information Engines workshop there for their valuable
feedback. J.P.C. acknowledges the kind hospitality of the Santa Fe Institute
and California Institute of Technology. This material is based on work
supported by, or in part by, the U.S. Army Research Laboratory and U.S. Army
Research Office under Grant No. W911NF-21-1-0048.

\bibliography{sceom}

\end{document}